\documentclass[%
 amsmath,amssymb,
 aps,
 twocolumn
]{revtex4-2}

\usepackage{graphicx}% Include figure files
\usepackage{dcolumn}% Align table columns on decimal point
\usepackage{bm}% bold math
\usepackage[colorlinks=true,citecolor=blue]{hyperref}% add hypertext capabilities
\usepackage{url}
\begin{document}

\title{Estimation of the pseudoscalar glueball mass based on a modified Transformer}

\author{Lin Gao}
\email{silvester\_gao@qq.com}

\affiliation{American Association for the Advancement of Science, Washington, DC 20005, USA}

\date{\today}% It is always \today, today,

\begin{abstract}
A modified Transformer model is introduced for estimating the mass of pseudoscalar glueball in lattice QCD. The model takes as input a sequence of floating-point numbers and produces a two-dimensional vector output. It integrates floating-point embeddings and positional encoding, and is trained using binary cross-entropy loss. The paper provides a detailed description of the model's components and training methods, and compares the performance of the traditional least squares method, the previously used deep neural network, and the modified Transformer in mass estimation. The results show that the modified Transformer model achieves smaller statistical uncertainty in mass estimation than the traditional least squares method. Additionally, compared to the deep neural network, this model utilizes positional encoding and can handle input sequences of varying lengths, offering enhanced adaptability. 
\end{abstract}

\maketitle

%\tableofcontents

\section{Introduction}
Gluons play a crucial role in the study of Quantum Chromodynamics (QCD) as they mediate the strong interaction\cite{GellMann1962,Gattringer2010}. In QCD theory, gluons carry color charge, and multiple gluons can interact to form relatively stable glueballs. Glueballs are classified by quantum numbers such as spin $J$, parity $P$, and charge conjugation $C$. The quantum state $J^{PC} = 0^{-+}$ corresponds to a pseudoscalar glueball, which is the focus of this study in estimating its mass.

This paper is an improvement upon the previous work titled "Study of the mass of pseudoscalar glueball with a deep neural network", continuing to use the negative tail $\widetilde{C}_q(r)$ of the correlation function of the topological charge density to extract the mass of the pseudoscalar glueball\cite{Gattringer2010,alexandrou2020comparison,gao2024mwgan}
\begin{equation}
\widetilde{C}_q(r) = A\frac{m}{4\pi^{2}r}K_{1}(m r),
\end{equation}
where \( m \) is the pseudoscalar glueball mass, \( A \) is an irrelevant normalization factor and $K_1$ is the modified Bessel function whose asymptotic form at large z is given by
\begin{equation}
K_1(z) \sim K_{1A}(z) = \sqrt{\frac{\pi}{2z}} \mathrm{e}^{-z} \left(1 + \frac{3}{8z}\right).
\end{equation}
For more details on the introduction, please refer to Ref.\cite{gao2024dnn}. Additionally, the data preparation about the lattice QCD in this paper is consistent with that in Ref.\cite{gao2024dnn}, so it will not be repeated here. This paper will focus primarily on the discussion of the improvements.

The main improvement in this paper is the use of a modified Transformer (MT) model to replace the previously used deep neural network (DNN). Transformer models have demonstrated outstanding performance in various sequence processing tasks. For instance, Chat Generative Pre-trained Transformer (ChatGPT), which is based on the Transformer model, excels in processing language sequences. In this study, I apply the MT architecture to process floating-point sequences of the negative tail $\widetilde{C}_q(r_n)$ of the correlation function. The input sequence length is $N \in [25,30]$, and the output is a two-dimensional vector related to factor $A$ and mass $m$. This setup is particularly well-suited for tasks involving the processing of precise numerical data. Moreover, the sequence of the negative values of the correlation function $\widetilde{C}_q(r_n)$, denoted as $c = (c_1, c_2, c_3, \dots, c_i, \dots, c_j, \dots, c_N)$, is used to estimate the mass of the pseudoscalar glueball. In this context, it is crucial to consider the position of each element within the sequence. In other words, if the positions of $c_i$ and $c_j$ are swapped, the resulting sequence used for mass estimation would differ from the original one. The previous DNN model did not incorporate positional encoding\cite{gao2024dnn}, but in this paper, positional encoding is introduced to explicitly account for the position of elements within the sequence. Additionally, when estimating the mass of the pseudoscalar glueball, the length of the correlation function sequence may not be fixed. The previous DNN model only handled sequences of fixed lengths. Therefore, the MT model is employed in this study to process input sequences of varying lengths.

The structure of this paper is as follows. First, the functionality of the MT model is introduced, followed by a discussion of the model's specific architecture and training details. Next, a comparison is made between the traditional least squares (LS) method, the DNN model, and the MT model. Finally, the paper presents the conclusions.

\section{Model functionality}

The MT model constructed in this paper has the following functionality. When the input is the negative portion of the correlation function, $c = (c_1, c_2, c_3, \dots, c_N)$, the two output parameters $P_1$ and $P_2$ are proportional to the factors $A$ and the mass $m$. It should be noted that $P_1$ and $P_2$ are constrained to the range $(0, 1)$. They are not directly equal to $A$ and $m$, but are related by proportional coefficients. This design choice is made because MT perform better when the output is within the range $(0, 1)$, a process known as normalization. The proportional coefficients will be provided later. Through this approach, the mass estimate $m$ can be obtained using the MT model.

To achieve the output described above, a neural network is constructed, which can be regarded as a functional $F$. After training, $F$ can map the following function
\begin{equation}
f(r_n, l_1, l_2) = -\frac{50 l_1 l_2}{\pi^2 a r_n} K_{1A}\left(\frac{2 l_2 r_n}{a}\right)
\end{equation}
to its parameters $l_1$ and $l_2$ approximately, where $l_1$ and $l_2$ are in the range $(0,1)$ and $K_{1A}(z)$ is the asymptotic form of the modified Bessel function at large z. Specifically, the input is a sequence consisting of the values of $f$ at different positions, and the output should approximately equal $l_1$ and $l_2$. It is important to note that the output is approximately equal to $l_1$ and $l_2$, rather than exactly equal to these two parameters, because the machine learning training is a process of gradual approximation, inherently involving some degree of approximation.

Specifically, when $l_1$ and $l_2$ are equal to $-A/100$ and $am/2$, respectively, we can obtain $f(r_n, -A/100, am/2) = \widetilde{C}_q(r_n)$. In other words, $\widetilde{C}_q(r_n)$ corresponds to a special case of $f(r_n, l_1, l_2)$. As previously mentioned, $l_1$ and $l_2$ should be within the range $(0,1)$, so $-A/100$ and $am/2$ should also be constrained within $(0,1)$. Based on known fitting results\cite{gao2024dnn,Athenodorou2020}, we can estimate a rough range for $m$, ensuring that $am/2$ falls within $(0,1)$. Similarly, we can ensure that $-A/100$ lies within $(0,1)$. Therefore, when the trained neural network can identify the parameters $l_1$ and $l_2$ of the function $f$, it can also identify the special case parameters $-A/100$ and $am/2$ of $\widetilde{C}_q$, thereby enabling an estimation of $m$.

\section{Model Architecture}

The MT model includes several key components, namely Input Embedding, Positional Encoding, Encoder-Decoder, Output Linear Layer, and Sigmoid Activation. Each of these components will be explained in detail below.

\subsection{Input Embedding}
The input sequence is denoted as $\mathbf{X} \in \mathbb{R}^{N_b \times N}$, where $N_b$ is the batch size and $N$ is a variable value $N \in [25,30]$. First, it is necessary to map each floating-point number in the input sequence to a high-dimensional vector for subsequent processing. A basic requirement for input embedding is that different floating-point numbers must be mapped to different high-dimensional vectors, otherwise, it would be impossible to distinguish between different values in later stages. Furthermore, since real numbers are continuous, it is not feasible to use $one-hot$ encoding to map them to a finite number of high-dimensional vectors. In this paper, the input embedding is achieved by projecting different floating-point numbers into distinct vectors in high-dimensional space using a linear layer
\begin{equation}
\mathbf{H}_{\text{input}} = \text{Linear}(\mathbf{X}).
\end{equation}
The dimension of the high-dimensional space selected in this paper, $d_{\text{model}}$, is 128. Using a linear transformation, this effectively maps different floating-point numbers to distinct high-dimensional vectors.

\subsection{Positional Encoding}
The sinusoidal positional encoding was introduced as an explicit solution in Ref.\cite{vaswani2017attention}. The formula for computing positional encoding is given as follows
\begin{equation}
\begin{split}
PE_{(k, 2i)} &= \sin\left({k}/{10000^{2i/d_{\text{model}}}}\right), \\
PE_{(k, 2i+1)} &= \cos\left({k}/{10000^{2i/d_{\text{model}}}}\right),
\end{split}
\end{equation}
where $d_{\text{model}}$ represents the dimension of the positional vector $\mathbf{PE}$. The components $PE_{(k, 2i)}$ and $PE_{(k, 2i+1)}$ correspond to the $2i$-th and $(2i+1)$-th elements of the encoding vector at position $k$, respectively.

\begin{figure}[htb]
    \centering
    \includegraphics[width=0.6\linewidth]{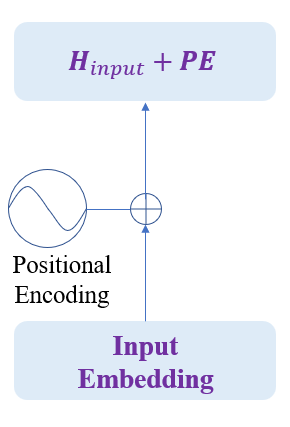}
    \caption{Positional Encoding.}
    \label{Positional_Encoding}
\end{figure}

The positional encoding layer provides the model with information about the relative positions of the inputs. As shown in Fig.~\ref{Positional_Encoding}, the input first goes through input embedding, followed by positional encoding using sinusoidal functions. The final encoded result, ${\mathbf{H}_{\text{input}} + \mathbf{PE}}$, is then fed into the encoder for further processing.

\subsection{Encoder-Decoder}
\begin{figure}[h]
    \centering
    \includegraphics[width=\linewidth]{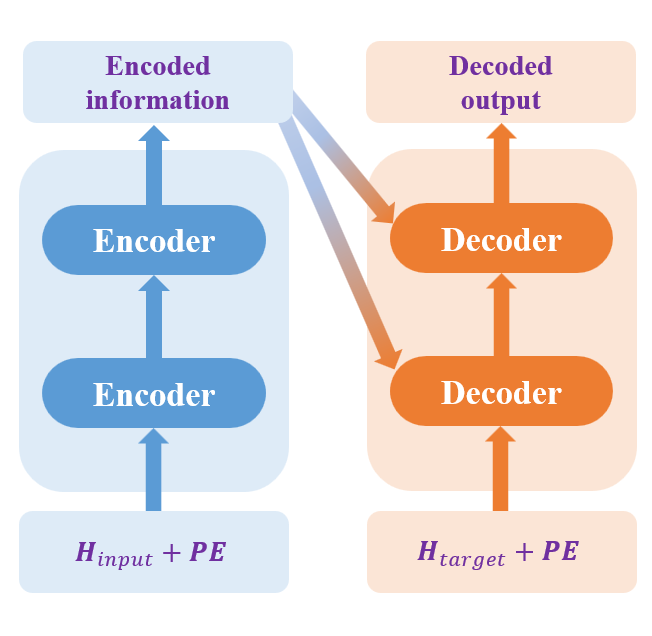}
    \caption{Encoder-Decoder}
    \label{Encorder_Decorder}
\end{figure}

The core of the MT model is the encoder-decoder structure. As shown in Fig.~\ref{Encorder_Decorder}, the encoder processes the input sequence with added positional encoding, while the decoder gradually generates the output sequence. The attention mechanism allows the model to dynamically focus on different parts of the input sequence.
\begin{equation}
\mathbf{H}_{\text{encoder}} = \text{Encoder}(\mathbf{H}_{\text{input}} + \mathbf{PE}),
\end{equation}
\begin{equation}
\mathbf{H}_{\text{decoder}} = \text{Decoder}(\mathbf{H}_{\text{encoder}}, \mathbf{H}_{\text{target}} + \mathbf{PE}).
\end{equation}

Self-attention with positional encoding is generally described by the following formula
\begin{equation}
\left\{\begin{aligned} 
\boldsymbol{q}_i =&\, (\boldsymbol{h}_i + \boldsymbol{p}_i)\boldsymbol{W}_Q \\ 
\boldsymbol{k}_j =&\, (\boldsymbol{h}_j + \boldsymbol{p}_j)\boldsymbol{W}_K \\ 
\boldsymbol{v}_j =&\, (\boldsymbol{h}_j + \boldsymbol{p}_j)\boldsymbol{W}_V \\ 
{\tilde a}_{l,j} =&\, softmax\left(\boldsymbol{q}_l \boldsymbol{k}_j^{\top }/\sqrt {d_k}\right)\\ 
\boldsymbol{o}_l =&\, \sum_j {\tilde a}_{l,j}\boldsymbol{v}_j 
\end{aligned}\right. ,
\end{equation}
where ${q}_i$ and ${q}_l$ are indexed differently to indicate that the number of ${q}_i$ and ${o}_l$ may not be the same, $\boldsymbol{h}_i$ represents the array obtained after the embedding step, and $\boldsymbol{p}_i$ is the positional encoding array. Both $\boldsymbol{h}_i$ and $\boldsymbol{p}_i$ are represented in component form, and the overall form corresponds to $ \mathbf{H}_{input}$ and $\mathbf{PE}$. Additionally, the MT model uses multi-head self-attention, with the number of heads set to 8.

\subsection{Output Linear Layer and Sigmoid Activation}
The decoder output is passed through a linear layer, followed by a Sigmoid activation function to generate the final output
\begin{equation}
\mathbf{Y} = \sigma(\text{Linear}(\mathbf{H}_{\text{decoder}})),
\end{equation}
where $\sigma$ denotes the Sigmoid function, which ensures that the output is within the range $(0, 1)$. The input dimension of the linear layer is $d_{\text{model}}$, and the output dimension is 2.

In the original Transformer paper\cite{vaswani2017attention}, the model uses softmax at the end, which produces a probability distribution. This is particularly useful for language tasks covered in the original paper. For instance, when inputting a text to a machine, the length of the model's output can vary depending on the content of the text. For example, in ChatGPT, the length of the response to a question can vary based on the question. However, in this paper, the goal is to find two parameters, which means the model's output length is fixed at 2. Therefore, it is unnecessary to use softmax to generate a probability distribution and produce outputs one by one, instead, the model can directly output the two parameters in a single step.

\section{Training}
The model is trained using binary cross-entropy loss
\begin{equation}
\mathcal{L} = -\frac{1}{N_b} \sum_{i=1}^{N_b} \left[ y_i \log(\hat{y}_i) + (1 - y_i) \log(1 - \hat{y}_i) \right],
\end{equation}
where $y_i$ and $\hat{y}_i$ represent the true labels and predicted labels, respectively. Since the model uses the Sigmoid function before outputting, the output $\hat{y}_i$ is within the range $(0, 1)$, thus avoiding $\log(0)$.

Additionally, the Adam optimizer is used during training with a learning rate of $3 \times 10^{-4}$\cite{kingma2014adam}. The model parameters are updated iteratively to minimize the loss function.

\section{Results and Discussion}

\begin{figure}[htb]
\includegraphics[width=\linewidth]{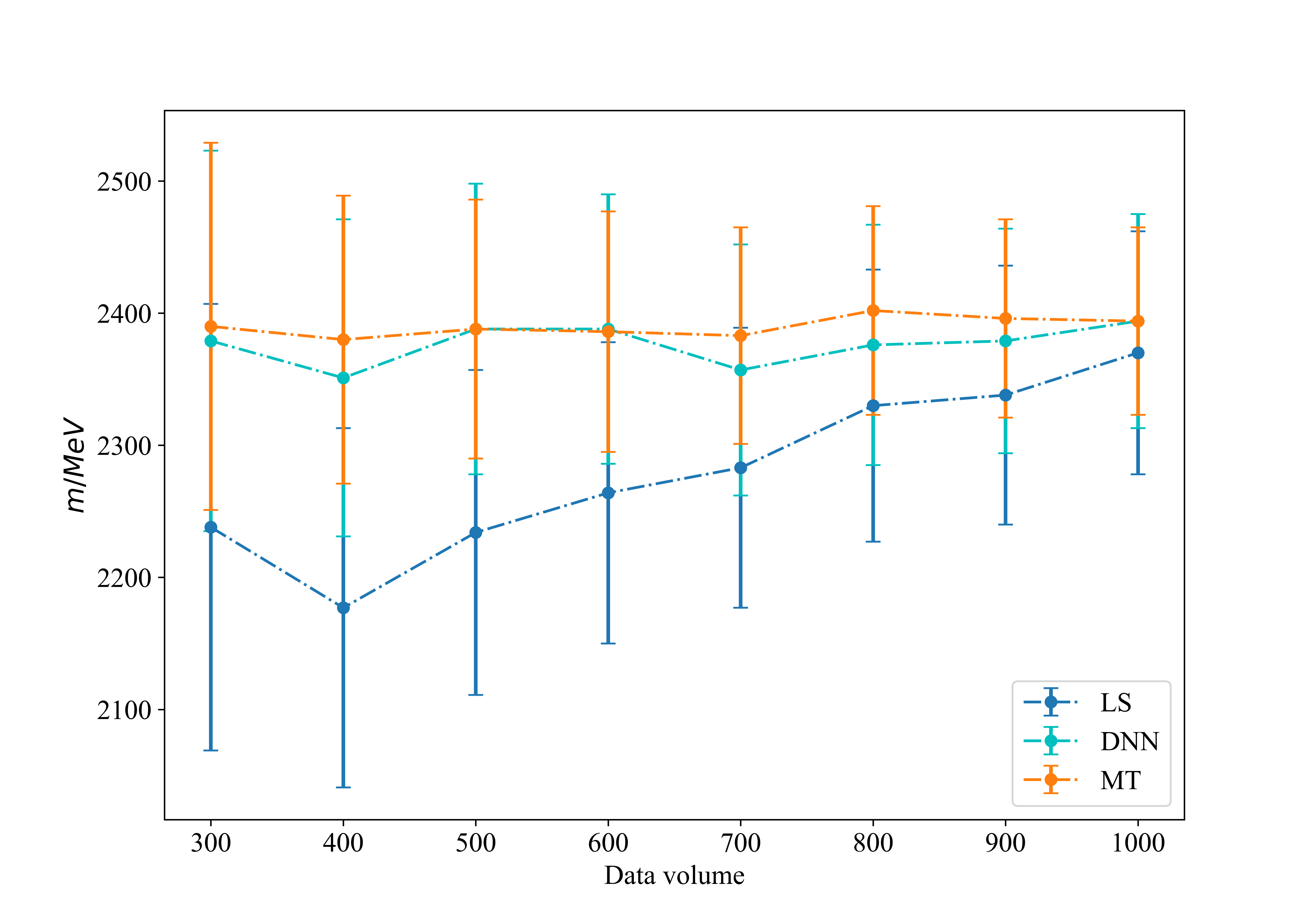}
\caption{\label{figure_LS_DNN_MT}The mass of the pseudoscalar glueball obtained using the traditional LS method and the machine learning approaches under different data volumes}
\end{figure}

First, the mass $m$ of the pseudoscalar glueball obtained using LS, DNN, and MT methods is compared across different configuration data volumes. As shown in Tab.~\ref{LS_ML_MT} and Fig.~\ref{figure_LS_DNN_MT}, the errors in the mass estimates decrease as the data volume increases for all three methods. Moreover, the two machine learning models, DNN and MT, exhibit smaller and more stable errors compared to the traditional LS method across various data sizes. Additionally, the MT model achieves higher precision in estimating the pseudoscalar glueball compared to the previously used DNN. It is necessary to calculate goodness-of-fit with $\chi^2$. For these three methods, the calculation shows that the values of $\chi^2$ per degree of freedom are all less than 2.5, indicating that three methods are generally appropriate in estimating the mass.

\begin{table}[htb]
\caption{\label{LS_ML_MT}The mass $m$ of the pseudoscalar glueball obtained using the traditional LS method, DNN and MT.}
\begin{ruledtabular}
\begin{tabular}{cccc}
\textrm{Data volume}&
\textrm{$m(LS)/MeV$}&
\textrm{$m(DNN)/MeV$}&
\textrm{$m (MT)/MeV$}\\
\colrule
500 & 2468(159) & 2530(125)  & 2559(112)  \\
600 & 2512(143) & 2512(114)  & 2537(103)  \\
700 & 2493(134) & 2524(104)  & 2545(97)  \\
800 & 2529(126) & 2544(96)  & 2551(89)  \\
900 & 2553(118) & 2522(92)  & 2556(84)  \\
1000 & 2612(112) & 2558(89)  & 2579(81)  \\
\end{tabular}
\end{ruledtabular}
\end{table}

\begin{figure}[htb]
\includegraphics[width=0.95\linewidth]{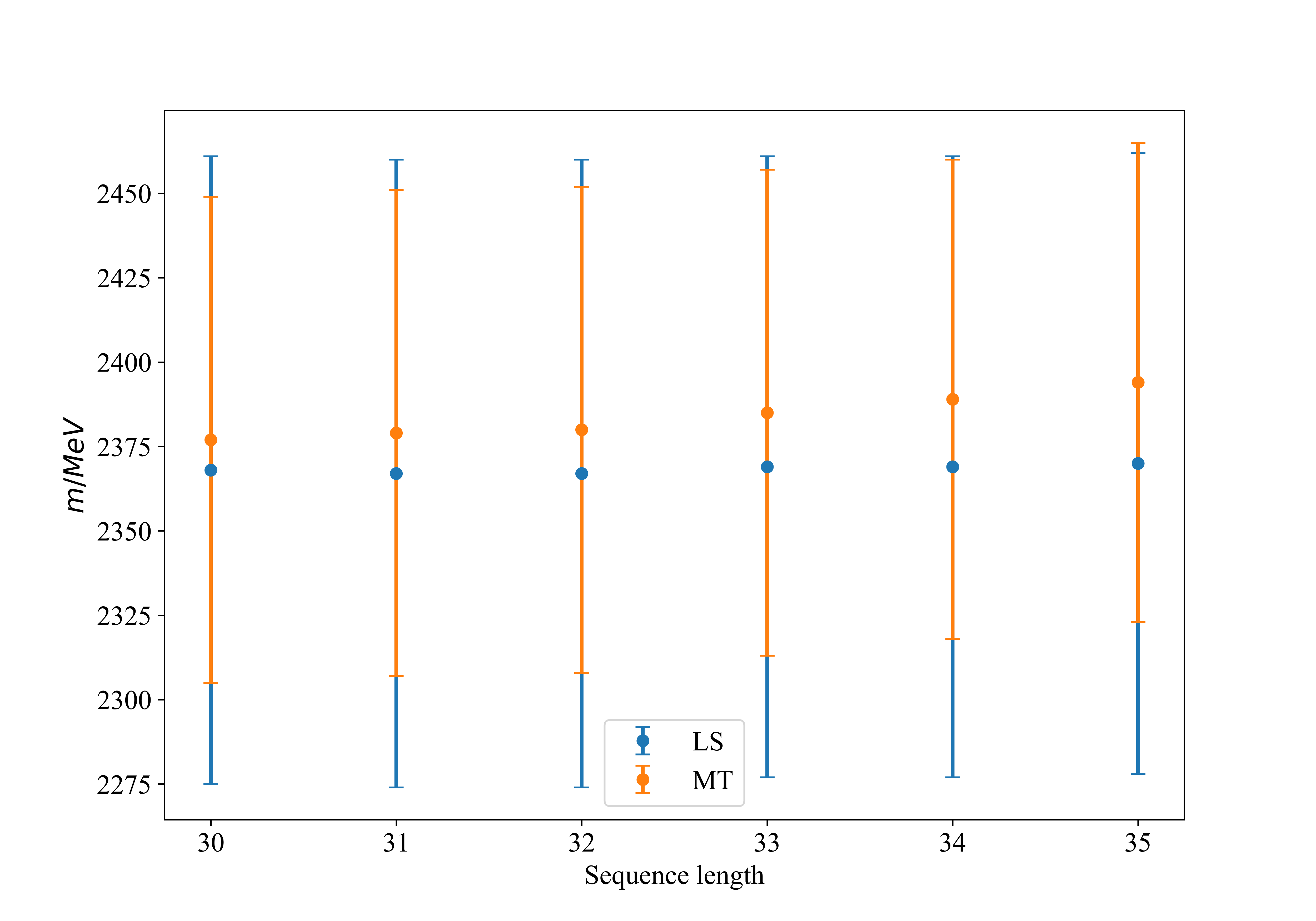}
\caption{\label{figure_m_length}The mass $m$ obtained by the traditional LS method and the MT model under different sequence length}
\end{figure}

Additionally, as shown in Fig.~\ref{figure_m_length}, with a data volume of 1000, I compared the mass of the pseudoscalar glueball obtained from $\widetilde{C}_q(r_n)$ sequences of different lengths. It can be observed that the mass values obtained by LS and MT do not vary significantly when the sequence length is between 25 and 30, and the statistical uncertainty of MT is smaller than that of the LS method.

\section{Conclusion}
This paper presents a modified Transformer (MT) model based on Transformer architecture for estimating the mass of the pseudoscalar glueball, using input sequences of $\widetilde{C}_q(r_n)$ composed of floating-point numbers.

To begin with, a linear layer is employed to implement input embedding for the floating-point sequence. Since real numbers are continuous and cannot be mapped to a finite number of high-dimensional vectors using one-hot encoding, this linear layer effectively maps different floating-point numbers to distinct high-dimensional vectors through linear transformation.

Moreover, sinusoidal positional encoding is used to encode the positions of elements in the input sequence. The model employs 2 encoders and 2 decoders for attention processing of the data, followed by a linear layer and a sigmoid function to normalize the output to the range (0,1). The model is trained using binary cross-entropy loss and the Adam optimizer.

Compared to the previously used DNN, the MT model offers improved precision in estimating the mass $m$ of the pseudoscalar glueball. Additionally, the MT model is adaptable to input sequences of varying lengths, enhancing its versatility. As the data volume increases, the errors in mass $m$ estimates from LS, DNN, and MT decrease progressively. Furthermore, both machine learning models, DNN and MT, exhibit smaller and more stable errors compared to the traditional LS method across different data sizes. For a data volume of 1000, the MT model estimates the mass $m$ as $2579(81) MeV$.

It is anticipated that the continued refinement of the MT model will further improve calculations in lattice QCD.

\textbf{Acknowledgments.}  This research utilized Chroma for simulating the lattice gauge field configurations\cite{Edwards_2005}. I extend my gratitude to the contributors of Chroma for their valuable contributions.

\bibliographystyle{apsrev4-2}
\bibliography{apssamp}

\end{document}